# The relationship between atmospheric stratification and internal wave processes.

**A.V. Kochin**
Central Aerological Observatory, Dolgoprudny, Russia.
amarl@mail.ru

**Abstract.**

The atmosphere is a resonant system and its oscillation spectrum is determined by the spatial distribution of parameters. For example, the Brunt-Väisälä frequency of internal gravity waves depends on the vertical temperature gradient. Therefore, the study of the spectra of internal wave processes can be used to estimate the spatial distribution of atmospheric parameters. The work is aimed at detecting wave fluctuations in the atmosphere and calculating atmospheric parameters based on the measured spectra. The rate of ascent of the radiosondes was used as a reference information, which was compared with the spectra of pressure fluctuations at the surface. The stratification parameters were calculated based on the Brunt-Väisälä frequency and showed good agreement with the upper-air sounding data.

## 1. Introduction

The atmosphere is in constant motion due to inhomogeneities of various origins and scales. These can be turbulent pulsations or vortex movements, as well as ordered flows of various scales. The atmosphere is a resonant system and external forces in the form of, for example, gradients of air flow velocities cause the appearance of their own internal oscillations. The study of the relationship between the resonance properties of a system and its structure has been studied in many works (Gossard E., Hooke W. 1975, Holton J. 2004, Khaykin S. and al 2015, Kochin A. 2023, Kshevetsky S., Kulichkov N. 2015). However, the problem cannot be considered fully investigated. The mechanisms of oscillation formation are formulated at the empirical level, similar to the problem of describing turbulence. Accordingly, the accumulation of experimental material on internal oscillations is an important stage in their research. Internal vibrations are determined by the physical nature of the returning force. The work investigated mesoscale processes in which the returning force can be either the elasticity of the air (acoustic vibrations) or the buoyancy force that forms vibrations called internal gravity waves (IGW). The resonance properties of such a system are determined by the spatial distribution of its parameters. Thus, the study of the resonance properties of the system makes it possible to obtain information about its structure, in particular, about the vertical temperature profile.

## 2. Internal gravity waves.

The creation of high-resolution barometers at the beginning of the last century led to the detection of a fluctuation in surface pressure with a period of about 10 minutes. The generally accepted explanation for this effect is the effect of buoyancy in the atmosphere stratified by the gravitational field and the formation of internal gravity waves (IGW). The IGW period is inversely proportional to the Brunt-Väisälä frequency (denoted by N in meteorological literature). For the IGW period, we have the following ratio

$$T_{BV} = \frac{1}{N} = 2\pi\sqrt{\frac{T}{g(\gamma_a - \gamma_r)}} \qquad (1)$$

Where T is the temperature °K, g is the acceleration of gravity, $\gamma a$ is the dry-adiabatic temperature gradient (1 deg/100 m), and $\gamma r$ is the real (observed) temperature gradient. Relation (1) relates the stratification of the atmosphere to its resonance properties.

Estimating stratification by vibration spectra requires specifying the type of vertical temperature profile. For this purpose, a standard atmosphere (ISA) is used in model problems in atmospheric physics. Up to an altitude of 80 km, the atmospheric parameters correspond to the average for a geographic latitude of 45°. For ISA, it is assumed that the temperature decreases vertically with an increase in altitude by 6.5 °C per 1 km to the level of 11 km (the conditional height of the tropopause) where the temperature becomes equal to −56.5 °C and almost ceases to change. Fig. 1 shows a view of the vertical temperature profile in ISA.

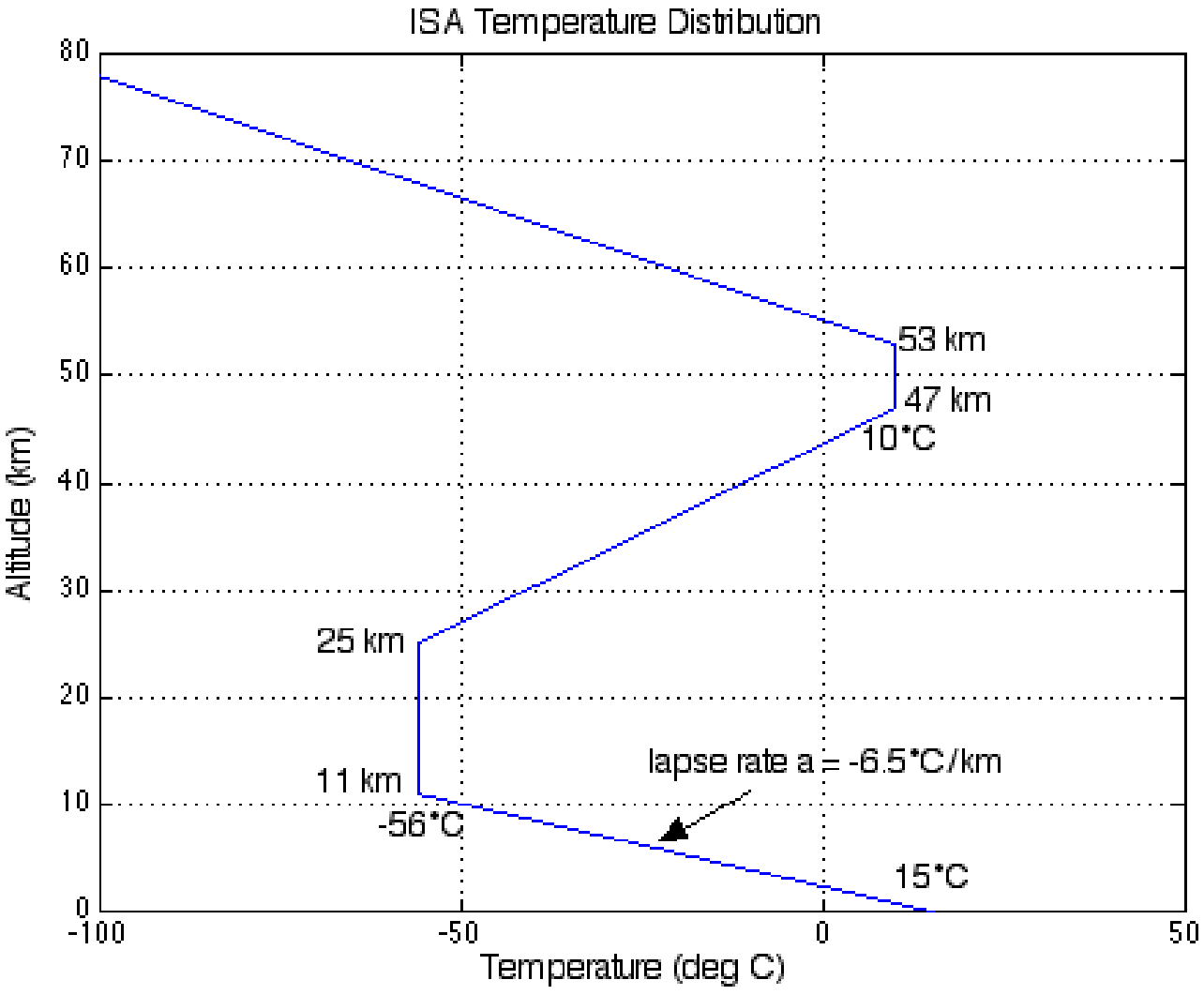

Fig. 1. Standard atmosphere temperature graph (ISA)
https://aviationperformance.wordpress.com/2020/08/29/basic-performance/.

The use of the ratio (1) and the vertical temperature profile in the standard atmosphere makes it possible to calculate the vertical profile of the Brunt-Väisälä frequency period, Fig. 2.

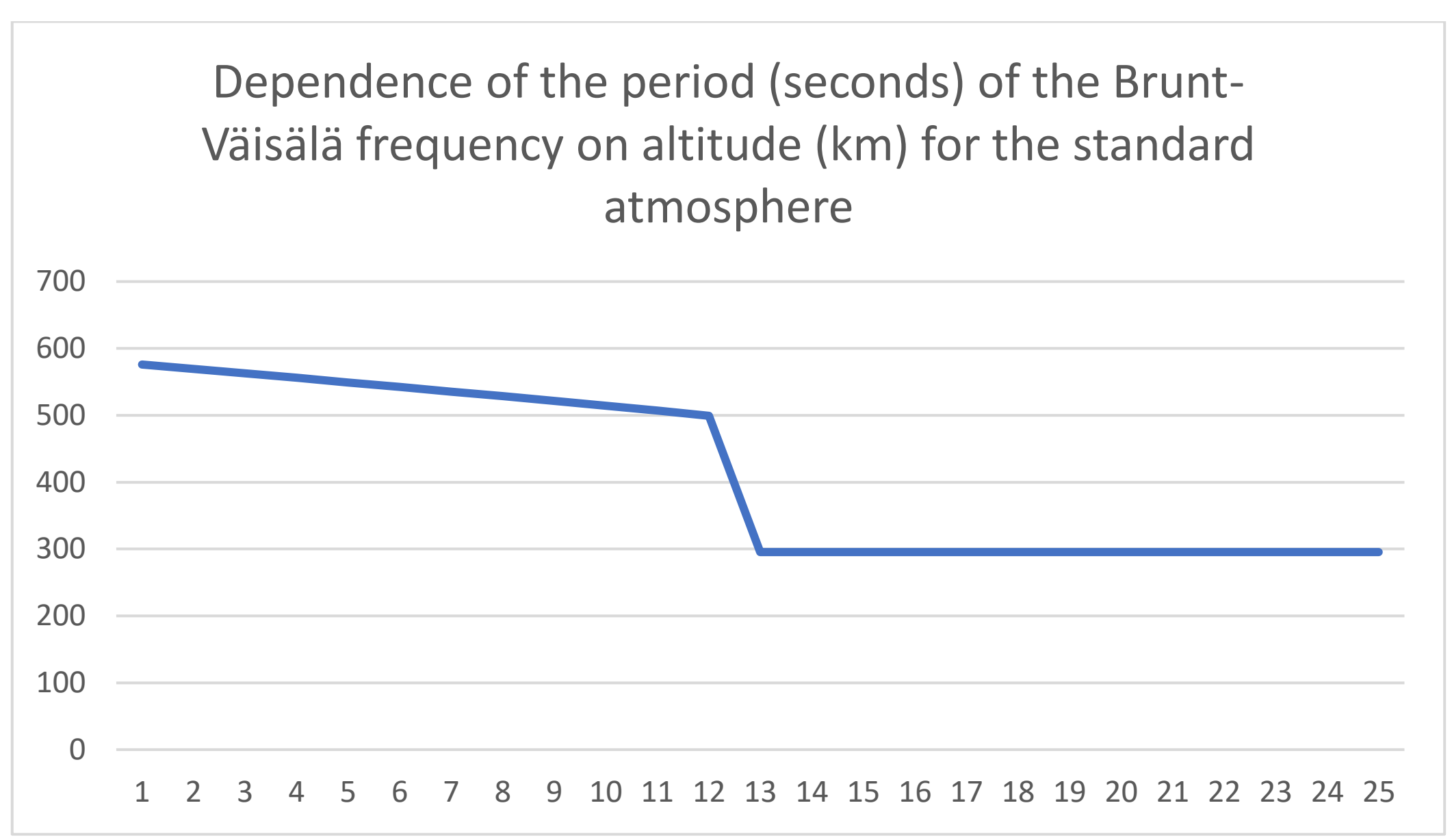

Fig.2. Vertical profile of the IGW period up to an altitude of 25 km.

This result allows us to make an assumption about the expected type of the IGW oscillation spectrum from the data of the microbarograph recording surface pressure (Fig. 3), which coincides with the results obtained back in the 1920s (Gossard E., Hooke W. 1975).

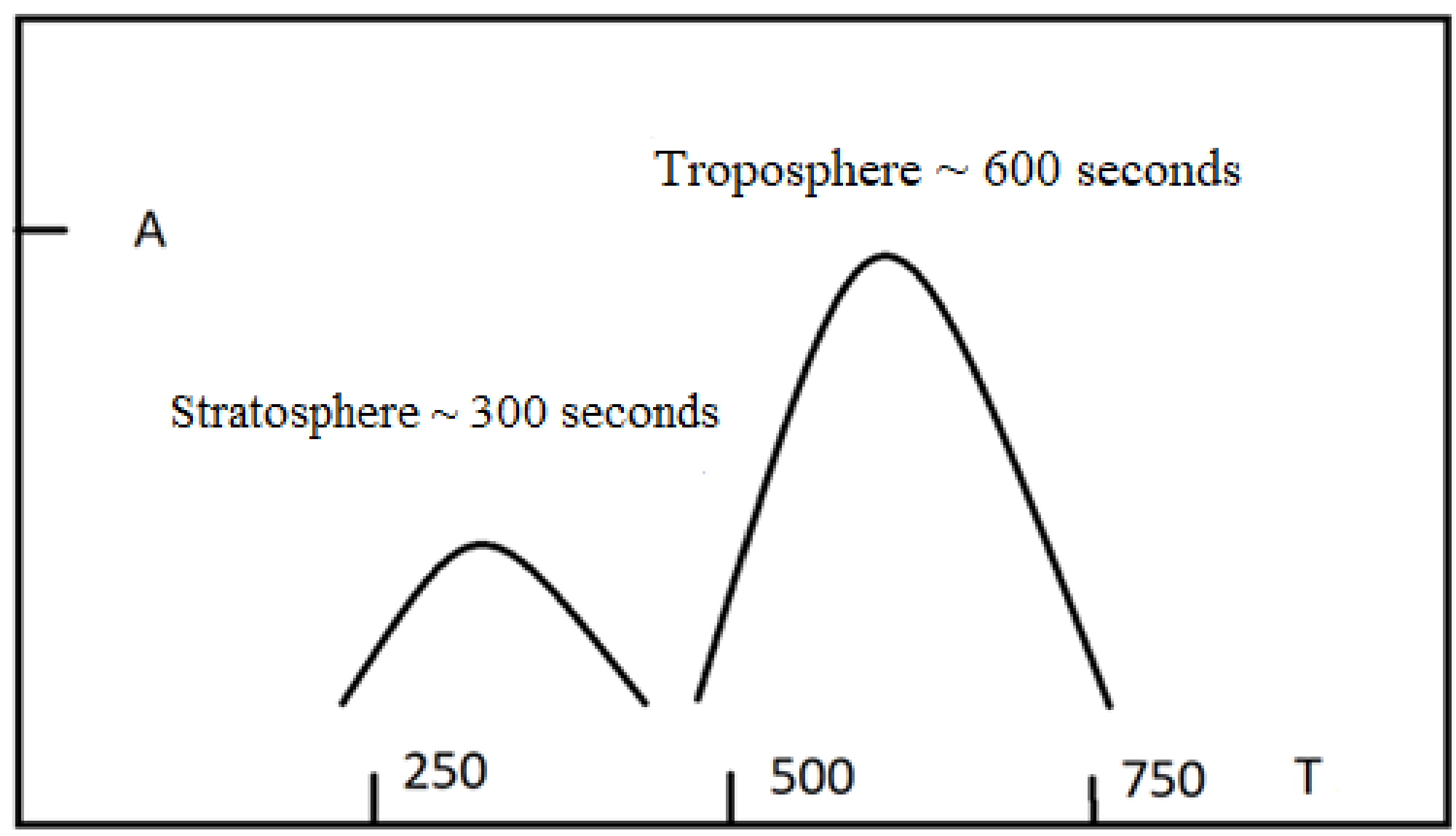

Fig.3. Conditional view of the spectra of the T oscillation periods in the stratosphere and troposphere in accordance with the ratio 1 for the standard atmosphere.

The assumption of a relatively small value of the stratospheric peak of 300 sec is associated with a smaller mass of air in the stratosphere and needs experimental confirmation.

**3. Experimental results.**

To detect internal fluctuations in the atmosphere, you need to measure the speed of the air. The most reliable method for measuring air velocity is the tracer method, which is the basis for wind speed measurements in upper-air sounding. To measure the vertical velocity component, it is necessary to take into account that the rate of rise of the radiosonde envelope varies due to internal causes with a characteristic period of 30 minutes or more. Speed fluctuations with a shorter period are usually caused by external factors and are almost always observed during radio sounding. Thus, the measurement of fluctuations in the rate of rise of the shell with a radiosonde is a direct method for measuring the parameters of atmospheric oscillations. A prerequisite is the use of high-precision sensing systems. To conduct the experiments, the sounding system of the MODEM company, France, based on the GPS navigation system, was used. Two radiosondes were launched into flight at intervals of 300 seconds and were accompanied by two ground stations. Fluctuations in vertical velocity reached 1 m/s, while the displacements of the radiosonde relative to the unperturbed trajectory (integral of the increment of velocity) were about one hundred meters, i.e. the atmosphere actually rises and falls by such a value due to oscillations. Especially informative is the difference in the velocities of radiosondes at the same altitude, which gives a clear spectrum with narrow spectral lines.

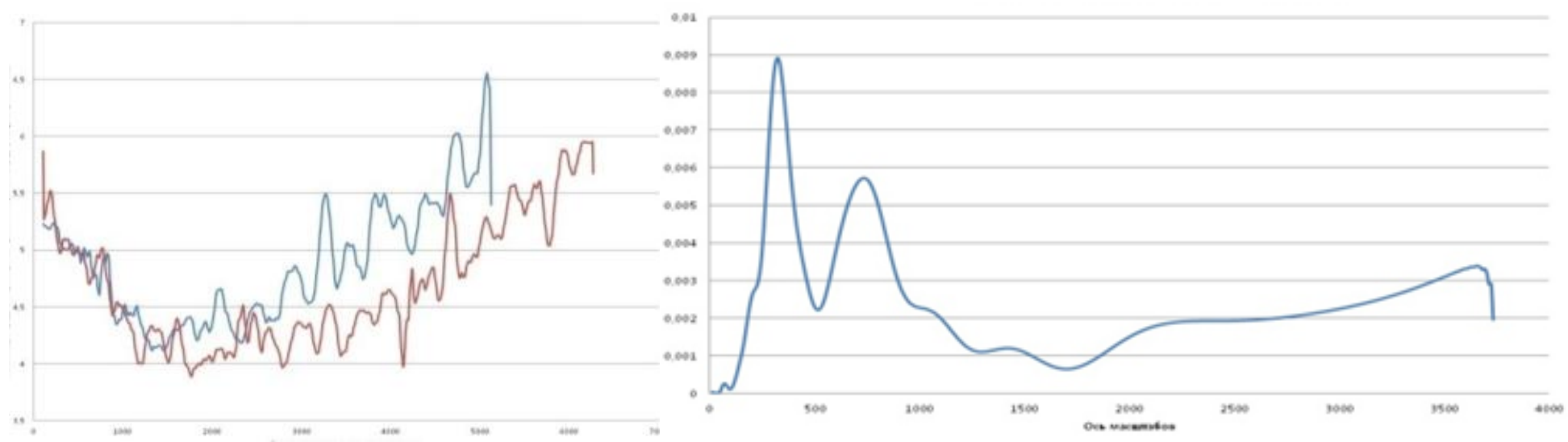

Fig. 4. Data of launch of two radiosondes with an interval of 300 seconds to indicate vertical movements by changing the speed of ascent of radiosondes. On the left is the speed of ascent of radiosondes m/s, on the right is the spectrum of the difference in the speed of ascent of radiosondes at the same altitude. The horizontal axis in the left figure represents the flight time in seconds, and the oscillation period in seconds in the right figure.

Experiments were also carried out to measure surface pressure and calculate the wavelet spectra from the measurement data. The results are shown in Fig. 5.

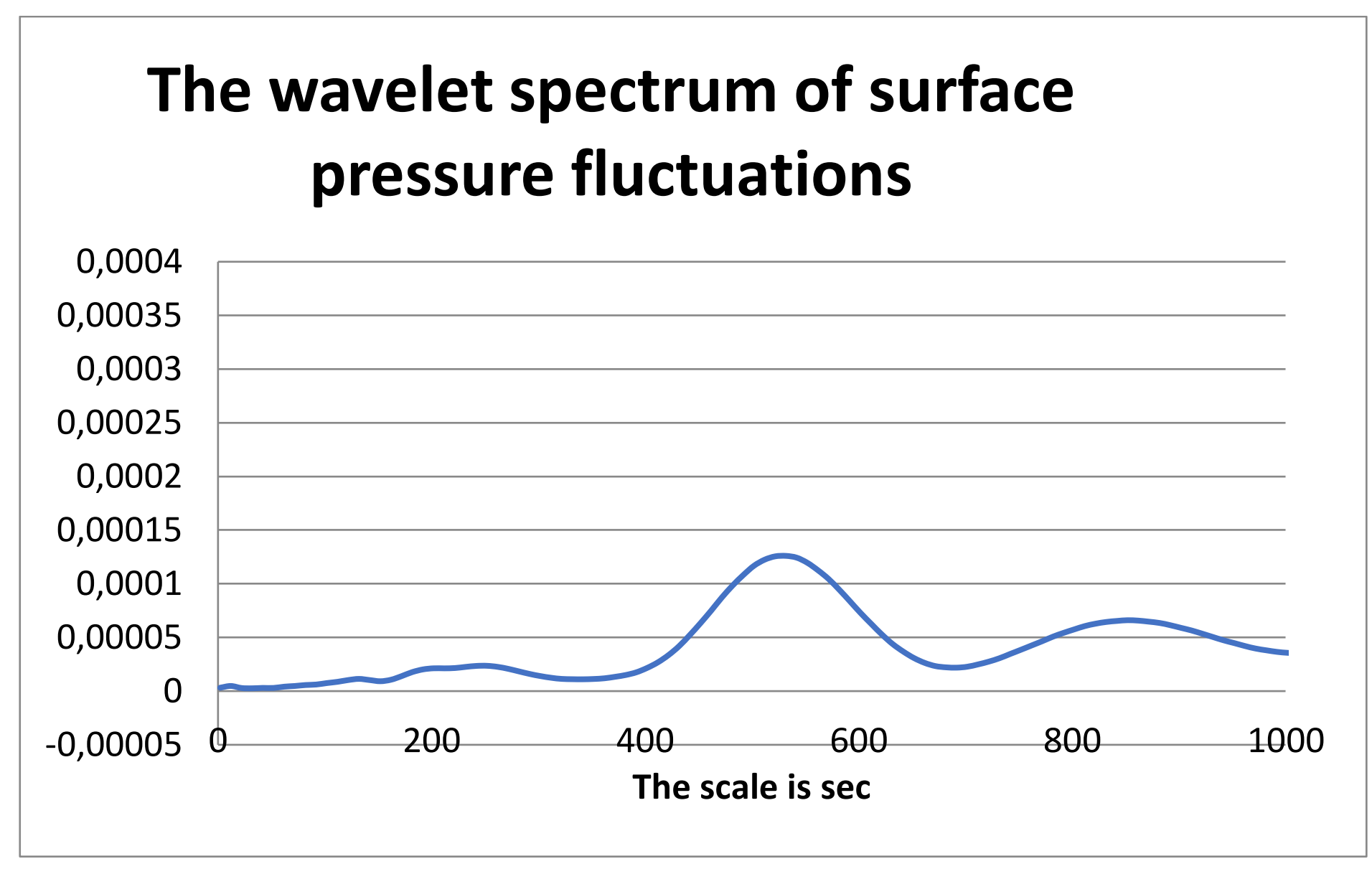

Fig. 5. Spectrum of surface pressure fluctuations.

It can be seen that the "tropospheric" period of 532 seconds in this case is determined satisfactorily, and the "stratospheric" period in the region of 300 seconds does not have a clear maximum.

**4. Assessment of the thermal characteristics of the atmosphere**

From the relation (1) for the IGW period, it is possible to obtain estimated ratios for the vertical temperature gradient $\gamma_T$ in the troposphere at the known surface temperature $T_E$

$$\gamma_T = \gamma_a - \frac{4\pi^2 T_E}{g T_{BV}^2} \qquad (2)$$

Where TE is the surface temperature °K, g is the acceleration due to gravity, $\gamma_a$ is the dry adiabatic temperature gradient (1 deg/100 m).

The relation for the stratosphere temperature assuming the isothermal temperature profile in it can be written as follows$T_S$

$$T_S = \frac{g T_{BV}^2 \gamma_a}{4\pi^2} \qquad (3)$$

Based on (2) and (3), a ratio for the tropopause height $H_{Tr}$ can be obtained

$$H_{Tr} = \frac{T_S - T_E}{\gamma_T} \qquad (4)$$

The above ratios make it possible, under certain assumptions, to estimate the stratification of the atmosphere, namely the vertical temperature gradient in the atmosphere and the temperature in the stratosphere, as well as the height of the tropopause. If we take the temperature in the equation for the IGW period to be equal to the surface temperature T from the weather station, it becomes possible to calculate the vertical temperature gradient in the troposphere from the central peak frequency in the range of 600 seconds. If we assume that the temperature gradient for the stratosphere is zero, then we can determine the temperature of the stratosphere from the central frequency of the peak in the range of 300 seconds.

For the twin launch, the measured temperature gradient in the troposphere was 0.8 °K/100 m at a surface temperature of 12°C. IGW period for such a gradient and temperature is 749 seconds. The maximum in the spectrum of oscillations of the velocity difference corresponds to the value of 735 seconds, which gives 0.79 °K/100 m. The observed temperature gradient in the stratosphere is variable from 0 to 0.09 °K/100 m, and the temperature in the tropopause is -55°C, above the tropopause in the stratosphere is - 47 °C. The IGW period for such conditions will vary from 293 to 298 seconds. The calculated value for the maximum in the spectrum of velocity difference oscillations is 325 seconds, which corresponds to a temperature of - 9°C, which coincides with the observed value according to radio sounding data only in an order of magnitude. The height of the tropopause on the radiosonde is about 10 km, which is also significantly different from the data on the IGW spectra.

It should be noted that the vertical temperature gradient actually observed from upper-air sounding data undergoes significant changes in altitude. Thus, the choice of the value of the vertical gradient to compare with the spectrum data is inevitably accompanied by noticeable uncertainty.

The measurement data of the microbarograph give very high uncertainty even for the troposphere and are practically inapplicable for determining the parameters of the stratosphere.

## 5. Conclusion

The main result of the work is the confirmation of the constant presence of oscillations with the Brunt-Väisälä frequency in the atmosphere. Since the period of the Brunt-Väisälä frequency is determined by atmospheric stratification, this makes it possible to estimate the parameters of atmospheric stratification.
The generally accepted method of detecting fluctuations is the analysis of the spectra of surface pressure fluctuations. To increase the reliability and accuracy of measurements, it is desirable to use additional sensors/sensors of a different type, such as an electric field strength sensor, and analyze the cross-correlation relationships between the sensors.
Detecting IGW on the Earth's surface will allow tracking changes in the synoptic situation, for example, specifying the time of passage of the atmospheric front. Information about the intensity of fluctuations in the atmosphere is also important for weather-dependent people.
The work was carried out with the help of colleagues from the Central Aerological Observatory, to whom the author expresses his gratitude.